\documentclass{elsart}

\usepackage{amssymb,graphicx}

\newcommand{\FOR}{Fortran~95}

\newcounter{bla}
\newenvironment{refnummer}{%
\list{[\arabic{bla}]}%
{\usecounter{bla}%
 \setlength{\itemindent}{0pt}%
 \setlength{\topsep}{0pt}%
 \setlength{\itemsep}{0pt}%
 \setlength{\labelsep}{2pt}%
 \setlength{\listparindent}{0pt}%
 \settowidth{\labelwidth}{[9]}%
 \setlength{\leftmargin}{\labelwidth}%
 \addtolength{\leftmargin}{\labelsep}%
 \setlength{\rightmargin}{0pt}}}
 {\endlist}

\begin{document}
\begin{frontmatter}


\title{New version announcement for TaylUR,
  an arbitrary-order diagonal automatic differentiation package for \FOR}

\author{G.M. von Hippel\thanksref{a}}
\thanks[a]{Corresponding author}
\address{Department of Physics, University of Regina, Regina,
  Saskatchewan, S4S 0A2, Canada}

\ead{vonhippg@uregina.ca}
\ead[url]{http://uregina.ca/\~\/vonhippg/}

% Abstract, keywords, PACS
\begin{abstract}
We present a new version of TaylUR, a \FOR\/ module to automatically
compute the numerical values of a complex-valued function's
derivatives with respect to several variables up to an arbitrary order
in each variable, but excluding mixed derivatives.
The new version fixes a potentially serious bug in the code for
exponential-related functions that could corrupt the imaginary parts
of derivatives, as well as being compatible with a wider range of
compilers.

\end{abstract}

\end{frontmatter}

% NEW VERSION PROGRAM SUMMARY.

{\bf NEW VERSION PROGRAM SUMMARY}

\begin{small}
\noindent
{\em Manuscript Title:} New version announcement for TaylUR,
an arbitrary-order diagonal automatic differentiation package
for Fortran 95                                                \\
{\em Authors:} G.M. von Hippel                                \\
{\em Program Title:} TaylUR                                   \\
{\em Journal Reference:}                                      \\
  %Leave blank, supplied by Elsevier.
{\em Catalogue identifier:}                                   \\
  %Leave blank, supplied by Elsevier.
{\em Licensing provisions:} none                              \\
{\em Programming language:} Fortran 95                        \\
{\em Computer:}  Any computer with a conforming Fortran 95
                 compiler                                     \\
{\em Operating system:} Any system with a conforming
                        Fortran 95 compiler                   
{\em Classification:} 4.12 Other Numerical Methods,
                      4.14 Utility                            \\
{\em Catalogue identifier of previous version:} ADXR\_v1\_0     \\
{\em Journal reference of previous version:}
Comput.~Phys.~Commun. {\bf 174} (2006) 569-576\\
{\em Does the new version supersede the previous version?:} yes \\

{\em Nature of problem:}\\
  Problems that require potentially high orders of derivatives with
  respect to some variables or derivatives of complex-valued
  functions, such as e.g. expansions of Feynman diagrams in particle
  masses in perturbative Quantum Field Theory. \\
   \\
{\em Solution method:}\\
  Arithmetic operators and Fortran intrinsics are overloaded to act
  correctly on objects of a defined type {\tt taylor}, which encodes a
  function along with its first few derivatives with respect to the
  user-defined independent variables. Derivatives of products and
  composite functions are computed using Leibniz's rule and F\`aa di
  Bruno's formula. \\
   \\
{\em Reasons for the new version:}\\
  The previous version [1] contained a potentially serious bug in the
  functions overloading the exponential-related intrinsics ({\tt EXP},
  {\tt LOG}, {\tt SIN}, {\tt COS}, {\tt TAN}, {\tt SINH}, {\tt COSH},
  {\tt TANH}), which could corrupt the imaginary parts of derivatives.
  It also contained some features which caused it to crash when
  compiled with certain compilers (notably the NAG and Lahey/Fujitsu
  compilers). \\ 
   \\
{\em Summary of revisions:}\\
  The bug in the exponential-related intrinsics has been corrected. A
  number of additional changes have been made to the code to enable
  better compatibility with a greater range of compilers, including
  the NAG and Lahey/Fujitsu compilers. Users of some of these
  compilers may have to define {\tt useintrinsic} as a preprocessor
  symbol when compiling TaylUR.\\
   \\
{\em Restrictions:}\\
  Memory and CPU time constraints may restrict the number of variables
  and Taylor expansion order that can be achieved. Loss of numerical
  accuracy due to cancellation may become an issue at very high
  orders. \\
   \\
{\em Unusual features:}\\
  No mixed higher-order derivatives are computed. The complex
  conjugation operation assumes all independent variables to be real. \\
   \\
{\em Running time:}\\
  The running time of TaylUR operations depends linearly on the number
  of variables. Its dependence on the Taylor expansion order varies
  from linear (for linear operations) through quadratic (for
  multiplication) to exponential (for elementary function calls). \\
   \\
{\em References:}
\begin{refnummer}
\item
  G.~M. von~Hippel,
  TaylUR, an arbitrary-order diagonal automatic differentiation package
  for Fortran 95,
  Comput.~Phys.~Commun. {\bf 174} (2006) 569-576.
\end{refnummer}

\end{small}

\end{document}